\documentclass{article}
\usepackage{graphicx}
\usepackage{amsmath}

\oddsidemargin  - 5 mm
\evensidemargin - 5 mm

\input{tcilatex}

\begin{document}

\title{Algebraic Geometry Approach in Theories with Extra Dimensions II.
Tensor Length Scale, Compactification and Rescaling \\
in Low-Energy Type I String Theory }
\author{Bogdan G. Dimitrov \thanks{%
Electronic mail: bogdan@theor.jinr.ru} \\
Bogoliubov Laboratory for Theoretical Physics\\
Joint Institute for Nuclear Research \\
6 Joliot - Curie str. \\
Dubna 141980, Russia}
\maketitle

\begin{abstract}
\ \ In this second part of the paper, dedicated to theories with extra
dimensions, a new physical notion about the "tensor length scale" is
introduced, based on the gravitational theories with covariant and
contravariant metric tensor components. Then the notion of
"compactification" in low energy type I string theory is supplemented by the
operation of "rescaling" of the contravariant metric components. For both
the cases of "rescaling+compactification" and "compactification+rescaling",
quasilinear differential equations in partial derivatives have been obtained
and the corresponding solutions have been found for the scale (length)
function and for the case of a flat $4D$ Minkowski space, embedded into a $%
5D $ space with an exponential warp factor. A differential equation has been
obtained and investigated also from the equality of the "rescaled" scalar
curvature with the usual one.
\end{abstract}

\section{INTRODUCTION}

In [1, 2], the algebraic geometry approach in gravity was developed, based
on the important distinction between covariant and contravariant metric
tensor components in the framework of the affine geometry approach [3, 4].

Also in a previous paper [5], a cubic algebraic equation was proposed, based
on the equivalence between the gravitational Lagrangian with the more
generally defined contravariant tensor and the usual Lagrangian.

In this paper, the same idea shall be exploited. The main difference will be
that instead of choosing the contravariant tensor in the form of the
factorized product $\widetilde{g}^{ij}=dX^{i}dX^{j}$ and solving the
corresponding algebraic equation with respect to $dX^{i}$, now this tensor
shall be chosen as $\widetilde{g}^{ij}=lg^{ij}$. Another difference in
comparison with the previous approach will be that instead of solving an
algebraic equation, this time the obtained equation will be considered as a
differential equation in partial derivatives, and it will be solved by means
of the \ known method of characteristics with respect to the function $l(%
\mathbf{x})$. This function is called the "length function" and as explained
in the previous part [6], it is a partial case of the newly introduced
notion of a "tensor length scale $l_{i}^{k}(\mathbf{x})$", which satisfies
the relation $g_{ij}\widetilde{g}^{ik}=l_{i}^{k}(\mathbf{x})$ in
gravitational theories with separately determined covariant $g_{ij}$ and
contravariant $\widetilde{g}^{ik}$ metric tensor components.

One of the main purposes of this paper is to find implementation of the
tensor length scale and function in the type I low-energy string theory. For
the purpose, the defined in [1,5] \textquotedblright
tilda\textquotedblright\ connection $\widetilde{\Gamma }_{ij}^{k}$ 
\begin{equation}
\widetilde{\Gamma }_{ij}^{k}\equiv \frac{1}{2}\widetilde{g}%
^{ks}(g_{js,i}+g_{is,j}-g_{ij,s})=l_{m}^{k}\Gamma _{ij}^{m}\text{ \ \ ,} 
\tag{1.1}
\end{equation}%
which turns out to be a linear combination of the Christoffell's connection
components $\Gamma _{ij}^{m}$ (and therefore is not independent from them),
is substituted in the known formulae for the "tilda" Riemann scalar
curvature 
\begin{equation}
\widetilde{R}=\widetilde{g}^{ij}\widetilde{R}_{ij}=\widetilde{g}^{ij}\left(
\partial _{k}\widetilde{\Gamma }_{ij}^{k}-\partial _{i}\widetilde{\Gamma }%
_{kj}^{k}+\widetilde{\Gamma }_{kl}^{k}\widetilde{\Gamma }_{ij}^{l}-%
\widetilde{\Gamma }_{ki}^{m}\widetilde{\Gamma }_{jm}^{k}\right) \text{ \ \ \
,}  \tag{1.2}
\end{equation}%
which constitutes the gravitational part of the string action. The
correctness of such a substitution is ensured by a known theorem from affine
geometry [3], the formulation of which and the proof for the partial $%
\widetilde{g}^{ij}=l\delta _{r}^{j}g^{ri}$ case have been presented in
Appendix A. \ 

Further, the coefficients in the type I low-energy string theory action are
compared before and after the compactification. Due to the newly introduced
operation of "rescaling", in this case not the standard simple algebraic
relations $M_{(4)}^{2}=\frac{(2\pi )^{7}}{V_{6}m_{s}^{4}g_{4}^{2}}$ and $%
\lambda =\frac{g_{4}^{2}V_{6}m_{s}^{6}}{(2\pi )^{7}}$ between the
four-dimensional Planck constant $M_{(4)}$, the string scale $m_{s}$ and the
electromagnetic coupling constant $g_{4}$ are obtained, but a \textit{%
quasilinear differential equation in partial derivatives for the length
function }$l(x)$. Moreover, since the compactification can be performed from
the "unrescaled" or from the "rescaled" gravitational part of the string
action, two such equations are obtained and the corresponding solutions by
the methods of characteristics have been derived. If one assumes that it is
irrelevant whether compactification or rescaling is performed at first, then
from the two differential equations a simple cubic algebraic equation can be
derived. From this equation, important inequalities for the parameters in
the low- energy type I string action can be obtained. \textit{Since the
length function }$l(x)$\textit{\ does not participate in these inequalities,
they might be relevent also for the present theories with extra dimensions.}

Another case, not related to type I string theory and to compactification
and for which the corresponding solutions have been obtained is the modified
gravitational action with the "tilda" connection (1.1) and again under the
choice $\widetilde{g}^{ij}=l_{r}^{j}g^{ri}=l\delta _{r}^{j}g^{ri}$ for the
contravariant components. For a flat $4D$ Minkowski space-time metric,
embedded in a $5D$ space of constant (or even non - constant) negative
curvature, the following expression for the length function $l(x)$ as a
solution of the quasilinear equation has been found: 
\begin{equation}
l^{2}=\frac{1}{1-const.\text{ }e^{24\text{ }k\text{ }\varepsilon \text{ }y}}%
\text{ \ \ \ \ \ \ }\varepsilon =\pm 1\text{ \ \ ,}  \tag{1.3}
\end{equation}%
From a physical point of view it is interesting to note that the
\textquotedblright scale function\textquotedblright\ will indeed be equal to
one (i.e. we have the usual gravitational theory with $\widetilde{g}%
^{ij}=g^{ij}$ and $l=1$) for $\varepsilon =-1$ and $y\rightarrow \infty $
(the s.c. infinite extra dimensions). However, for $\varepsilon =+1$ there
will be even a decrease of the \textquotedblright length
function\textquotedblright\ due to the exponential factor in the denominator.

\section{\protect\bigskip TENSOR \ LENGTH \ SCALE, RESCALING \ AND \
COMPACTIFICATION \ IN \ THE \ LOW \ ENERGY \ ACTION \ OF \ TYPE \ I \ TEN -
DIMENSIONAL \ STRING \ THEORY}

Now an \ example of the possible application of theories with covariant and
contravariant metrics shall be given, related to the low - energy action of
type I string theory in ten dimensions [7, 8, 9, 10] 
\begin{equation}
S=\int d^{10}x\left( \frac{m_{s}^{8}}{(2\pi )^{7}\lambda ^{2}}R+\frac{1}{4}%
\frac{m_{s}^{6}}{(2\pi )^{7}\lambda }F^{2}+...\right) \text{ }=\int
d^{4}xV_{6}(......)\text{\ \ \ \ ,}  \tag{2.1}
\end{equation}%
where $\lambda \sim \exp (\Phi )$ is the string coupling (as remarked in
[10], in the first term the coupling is $\lambda ^{2}$, because it is
generated by an world - sheet path integral on an sphere and the coupling $%
\lambda $ in the second term - by an world - sheet path integral on the
disc), $m_{s}$ is the string scale, which we can identify with $m_{grav.}$.
Compactifying to $4$ dimensions on a manifold of volume $V_{6}$, one can
identify the resulting coefficients in front of the $R$ and $\frac{1}{4}%
F^{2} $ terms with $M_{(4)}^{2}$ and $\frac{1}{g_{4}^{2}}$, from where one
obtains [7] 
\begin{equation}
M_{(4)}^{2}=\frac{(2\pi )^{7}}{V_{6}m_{s}^{4}g_{4}^{2}}\text{ \ \ \ ; \ \ \
\ \ }\lambda =\frac{g_{4}^{2}V_{6}m_{s}^{6}}{(2\pi )^{7}}\text{ \ \ \ .} 
\tag{2.2}
\end{equation}%
The physical meaning of the performed identification is that since the
length scale $\sqrt{\alpha ^{^{\prime }}}$ of string theory, the volume $V$
of the (Calabi - Yau) manifold and the expectation value of the dilaton
field cannot be determined experimentally, they can be adjusted in such a
way so that to give the desired values of the Newton's constant, the GUT
(Grand Unified Theory) scale $M_{GUT}$\ and the GUT\ coupling constant [10].
It should be stressed that in the weakly coupled heterotic string theory
(when there are no different string couplings $\lambda \sim \exp (2\Phi )$
and $\lambda \sim \exp (\Phi )$, but just one), the obtained bound on the
Newton's constant [10] $G_{N}\geq \frac{\alpha ^{\frac{4}{3}_{GUT}}}{%
M_{GUT}^{2}}$ is too large, but in the same paper [10] it was remarked that
\textquotedblright the problem might be ameliorated by considering an
anisotropic Calabi - Yau with a scale $\sqrt{\alpha ^{^{\prime }}}$\ in $d$\
directions and $\frac{1}{M_{GUT}}$\ in $(6-d)$\ directions\textquotedblright
.\textbf{\ }

Now we shall propose, in the spirit of the affine geometry approach, how
such a different metric scale on the given manifold can be introduced by
defining more general contravariant tensors. The key idea is that the
contraction of the covariant metric tensor $g_{ij}$ with the contravariant
one $\widetilde{g}^{jk}=dX^{j}dX^{k}$ gives exactly (when $i=k$) the length
interval [5] 
\begin{equation}
l=ds^{2}=g_{ij}dX^{j}dX^{i}\text{ \ \ .}  \tag{2.3}
\end{equation}%
Naturally, for $i\neq k$ the contraction will give a tensor function $%
l_{i}^{k}=g_{ij}dX^{j}dX^{k}$, which can be interpreted as a
\textquotedblright tensor\textquotedblright\ length scale for the different
directions. In the spirit of the remark in [10], one can take for example 
\begin{equation}
l_{i}^{k}=g_{ij}dX^{j}dX^{k}=L_{1}\delta _{i}^{k}\text{ \ for \ }%
i,j,k=1,....,d\text{ \ \ \ \ ,}  \tag{2.4}
\end{equation}%
\begin{equation}
l_{a}^{b}=g_{ac}dX^{c}dX^{b}=L_{2}\delta _{a}^{b}\text{ \ for \ }%
a,b,c=1,....,6-d\text{ \ \ \ \ .}  \tag{2.5}
\end{equation}%
For simplicity and as a starting point, further we shall assume that for all
indices $i,j,k...$ 
\begin{equation}
l_{i}^{k}=l\text{ }\delta _{i}^{k}\text{ \ \ .}  \tag{2.6}
\end{equation}%
In fact, this will be fulfilled if we assume that the contravariant metric
tensor components $\widetilde{g}^{ij}$ are proportional to the usual inverse
contravariant metric tensor $g^{ij}$ with a function of proportionality $l(%
\mathbf{x})$, i.e. $\widetilde{g}^{ij}=l(\mathbf{x})g^{ij}$ (it will be
called a \textquotedblright conformal\textquotedblright\ rescaling). Further
we shall call the function $l(\mathbf{x})$ \textit{\textquotedblright a
length scale function}\textquotedblright .

Our next purpose will be to prove that if one imposes the requirement for
invariance of the low - energy type I string action (2.1) under the
\textquotedblright conformal\textquotedblright\ rescaling, i.e. 
\begin{equation*}
S=\int d^{10}x\left( \frac{m_{s}^{8}}{(2\pi )^{7}\lambda ^{2}}\widetilde{R}+%
\frac{1}{4}\frac{m_{s}^{6}}{(2\pi )^{7}\lambda }\widetilde{F}^{2}\right)
=\int d^{4}xV_{6}\left( ....\right) =
\end{equation*}%
\begin{equation}
=\int d^{4}x\left( M_{(4)}^{2}R+\frac{1}{4}\frac{1}{g_{4}^{2}}F^{2}\right) 
\text{ \ \ \ \ ,}  \tag{2.7}
\end{equation}%
\ \ then the length scale $l(x)$\ will be possible to be determined from a
differential equation in partial derivatives.\textbf{\ }In other words,
unlike the previously described in [7, 8, 9, 10] case, when the coefficients
in front of $R$ and $F^{2}$ before and after the compactification are
identified, here we shall propose another approach to the same problem.
Concretely, first a rescaling of the contravariant metric components shall
be performed, and after that the compactification shall be realized,
resulting again in the R.H.S. of the standard $4D$ action (2.1).

However, in principle another approach is also possible. One may start from
the \textquotedblright unrescaled\textquotedblright\ ten - dimensional
action (2.1), then perform a compactification to the four - dimensional
manifold and afterwards\textbf{\ }a transition to the usual
\textquotedblright rescaled\textquotedblright\ scalar quantities $\widetilde{%
R}$ and $\widetilde{F}^{2}$. Then it is required that the \textquotedblright
unrescaled\textquotedblright\ ten - dimensional effective action (2.1) (i.e.
the L. H. S. of (2.1)) is equivalent to the four - dimensional effective
action after compactification, but in terms of the rescaled quantities $%
\widetilde{R}$ and $\widetilde{F}^{2}$ in the R. H. S of (2.1). This can be
expressed as follows \ 
\begin{equation*}
S=\int d^{10}x\left( \frac{m_{s}^{8}}{(2\pi )^{7}\lambda ^{2}}R+\frac{1}{4}%
\frac{m_{s}^{6}}{(2\pi )^{7}\lambda }F^{2}\right) =\int d^{4}xV_{6}\left(
....\right) =
\end{equation*}%
\begin{equation}
=\int d^{4}x\left( M_{(4)}^{2}\widetilde{R}+\frac{1}{4}\frac{1}{g_{4}^{2}}%
\widetilde{F}^{2}\right) \text{ \ \ \ \ .}  \tag{2.8}
\end{equation}%
In the next subsections both cases shall be investigated, deriving the
corresponding (quasilinear) differential equations in partial derivatives
and moreover, finding concrete solutions of these equations for the special
case of the metric of a flat $4D$ Minkowski space, embedded in a five -
dimensional $ADS$ space of constant negative curvature. It will be shown
also that for a definite scale factor $h(y)=\beta y^{n}$ ($\beta $ is a
constant) in front of the extra -coordinate $y$ in the metric, the derived
differential equations are still solvable, in spite of the fact that the
five - dimensional space is no longer of a constant negative curvature.
Besides the opportunity to extend the results to such spaces of non-constant
curvature, there is one more reason for the necessity to investigate such
quasilinear differential equations for concrete cases - in principle,
examples can be given, when such equations cannot be explicitely solved. But
evidently, some special kinds of chosen metrics will allow the solution of
these equations and consequently the determination of the scale length
function $l(x)$\ in terms of all the important parameters in the low -
energy type I string thery action. If for certain metrics this is possible ,
then it will turn out to be possible to test whether there will be
deviations from the standardly known gravitational theory with $l=1$, if the
electromagnetic coupling constant $g_{4}$, the $4D$\ Planck constant $%
M_{(4)} $, the string scale $m_{s}$\ and the string coupling $\lambda $\ are
known, presumably from future experiments or cosmological data. Even if one
assumes that there no deviations from the standard theory with $l=1$, the
obtained solutions will allow to find some new relations between the above
mentioned parameters. The obtained differential equations in the limit of $%
l=1$ will result in the simple algebraic relations (2.2), already found in
the literature.

One may also require the equivalence of the two approaches, expressed
mathematically by (2.7) and (2.8), although for the moment it is not known
whether there is some physical reason for this equivalence.

\section{\protect\bigskip ALGEBRAIC RELATION \ AND A \ QUASILINEAR \
DIFFERENTIAL \ EQUATION \ IN \ PARTIAL \ DERIVATIVES \ FROM \ THE \
''RESCALED + COMPACTIFIED'' \ LOW - ENERGY TYPE I \ STRING \ ACTION}

\bigskip In order to rewrite the \textquotedblright
rescaled+compactified\textquotedblright\ string action (2.7), let us first
define the \textquotedblright rescaled\textquotedblright\ square of the
electromagnetic field strength as 
\begin{equation*}
\widetilde{F}^{2}=\widetilde{F}_{AB}\widetilde{F}^{AB}=F_{AB}\widetilde{g}%
^{AM}\widetilde{g}^{BN}F_{MN}=
\end{equation*}%
\begin{equation}
=l^{2}F_{AB}g^{AM}g^{BN}F_{MN}=l^{2}F^{2}\text{ \ \ \ .}  \tag{3.1}
\end{equation}%
Using the formulaes for the Riemann tensor and for the rescaled affine
connection 
\begin{equation}
\widetilde{\Gamma }_{AC}^{D}=\widetilde{g}^{DG}g_{GF}\Gamma
_{AC}^{F}=l\Gamma _{AC}^{D}\text{ \ \ ,}  \tag{3.2}
\end{equation}%
the rescaled scalar gravitational curvature $\widetilde{R}$ can be written
as 
\begin{equation*}
\widetilde{R}=\widetilde{g}^{DG}\widetilde{g}_{GF}\widetilde{R}_{ABCD}=\frac{%
1}{2}l^{2}g^{AC}g^{BD}(g_{AD,BC}+g_{BC,AD}-g_{AC,BD}-
\end{equation*}%
\begin{equation}
-g_{BD,AC})+l^{4}g^{AC}g^{BD}g_{FG}(\Gamma _{CB}^{F}\Gamma _{AD}^{G}-\Gamma
_{DB}^{F}\Gamma _{AC}^{G})=  \tag{3.3}
\end{equation}%
\begin{equation}
=l^{4}R-\frac{1}{2}%
l^{2}(l^{2}-1)g^{AC}g^{BD}(g_{AD,BC}+g_{BC,AD}-g_{AC,BD}-g_{BD,AC})\text{ \
\ \ \ .}  \tag{3.4}
\end{equation}%
Substituting the above expressions (3.1)\ and (3.4) for $\widetilde{F}^{2}$
and $\widetilde{R}$ into the L. H. S. of the low - energy string action
(2.7) and setting up equal the corresponding coefficients in front of the $%
\frac{1}{4}F^{2}$ term in the L. H. S. and the R. H. S. of (2.7), one can
derive 
\begin{equation}
\lambda =\frac{g_{4}m_{s}^{6}V_{6}}{(2\pi )^{7}}l^{2}\text{ \ \ \ .} 
\tag{3.5}
\end{equation}%
This is almost the same expression as in (2.2), but now corrected wih the
function of proportionality $l^{2}(\mathbf{x})$. The string coupling $%
\lambda $ is thus a non - local physical quantity, depending on the space -
time coordinates.

Next, after the elimination of the terms with $\frac{1}{4}F^{2}$ on both
sides of (2.7)\ and substituting the found formulae for $\lambda $ into the
resulting expression on both sides of (2.7), one derives the algebraic
relation 
\begin{equation}
\left[ \frac{(2\pi )^{7}}{V_{6}m_{s}^{4}g_{4}^{4}}-M_{(4)}^{2}\right] R=%
\frac{(2\pi )^{7}(l^{2}-1)}{2m_{s}^{4}V_{6}l^{2}g_{4}^{2}}g^{AC}g^{BD}(....)%
\text{ \ \ .}  \tag{3.6}
\end{equation}%
For brevity, the brackets $(....)$ will denote the term in (3.4)\ with the
second derivatives of the metric tensor. For $l=1$, as expected, we obtain
the usual relation for $M_{(4)}^{2}$ as in (2.2). Therefore, physically any
possible deviations from relation (2.2) can be attributed to the appearence
of the new length scale $l(\mathbf{x})$. Let us introduce the notation 
\begin{equation}
\beta \equiv \left[ \frac{(2\pi )^{7}}{V_{6}m_{s}^{4}g_{4}^{4}}-M_{(4)}^{2}%
\right] m_{s}^{4}V_{6}\frac{2}{(2\pi )^{7}}  \tag{3.7}
\end{equation}%
and assume that the deviation from the relation $M_{(4)}^{2}=\frac{(2\pi
)^{7}}{V_{6}m_{s}^{4}g_{4}^{2}}$ is small, i.e. $\beta \ll 1$.Then the
length scale $l(x)$ can be expressed from the algebraic relation (3.6) as 
\begin{equation}
l^{2}=\frac{1}{1-\beta \frac{R}{g^{AC}g^{BD}(...)}}\approx 1+\beta \frac{R}{%
g^{AC}g^{BD}(...)}\text{ \ \ \ .}  \tag{3.8}
\end{equation}%
\textit{Consequently the deviation from the \textquotedblright
standard\textquotedblright\ length scale }$l=1$\textit{\ in the case of a
gravitational theory with }$l\neq 1$\textit{\ in the case of small }$\beta $%
\textit{\ shall be proportional to the ratio }$\frac{R}{g^{AC}g^{BD}(...)}$%
\textit{.} In the concrete example of an $4D$ Minkowski space, embedded in a 
$5D$ $ADS$ space of constant negative curvature, this ratio will be 
\begin{equation}
\frac{R}{g^{AC}g^{BD}(...)}=\frac{(-8k^{2})}{(-32k^{2})}=\frac{1}{4}\text{ \ 
}  \tag{3.9}
\end{equation}%
and therefore, this constant factor will not affect the smallness of the
number $\beta \frac{R}{g^{AC}g^{BD}(...)}$.\textit{The above result has also
an important physical meaning - the zero value of the number }$\beta $%
\textit{\ (which signifies the fulfillment of the relation }$M_{(4)}^{2}=%
\frac{(2\pi )^{7}}{V_{6}m_{s}^{4}g_{4}^{2}}$\textit{) is directly connected
with the usual length scale }$l=1$\textit{\ in gravity theory.}

Let us now derive the differential equation in partial derivatives, starting
from the second representation of the \textquotedblright
rescaled\textquotedblright\ scalar gravitational curvature $\widetilde{R}$ $%
\ $by means of the \textquotedblright rescaled\textquotedblright\ Ricci
tensor $\widetilde{R}_{ij}$ 
\begin{equation}
\widetilde{R}=\widetilde{g}^{AB}\widetilde{R}_{AB}=lg^{AB}\left[ \frac{%
\partial \widetilde{\Gamma }_{AB}^{C}}{\partial x^{C}}-\frac{\partial 
\widetilde{\Gamma }_{AC}^{C}}{\partial x^{B}}+\widetilde{\Gamma }_{AB}^{C}%
\widetilde{\Gamma }_{CD}^{D}-\widetilde{\Gamma }_{AC}^{D}\widetilde{\Gamma }%
_{BD}^{C}\right] \text{ \ \ \ .}  \tag{3.10}
\end{equation}%
It can easily be found that the rescaled gravitational curvature is
expressed through the usual one as 
\begin{equation*}
\widetilde{R}=\widetilde{g}^{AB}\widetilde{R}_{AB}=l^{2}R+l^{2}(l-1)g^{AB}%
\left( \Gamma _{AB}^{C}\Gamma _{CD}^{D}-\Gamma _{AC}^{D}\Gamma
_{BD}^{C}\right) +
\end{equation*}%
\begin{equation}
+l\text{ }\frac{\partial l}{\partial x^{C}}g^{AB}\Gamma _{AB}^{C}-l\frac{%
\partial l}{\partial x^{B}}g^{AB}\Gamma _{AC}^{C}\text{ \ \ \ .}  \tag{3.11}
\end{equation}%
Again, this expression and also (2.9) for $\widetilde{F}^{2}$ are
substituted into the L. H. S. of the action (2.7) and the corresponding
coefficients in front of the term $\frac{1}{4}F^{2}$ in the L. H. S. and the
R. H. S. of (2.2)\ are set up equal. Thus one obtains 
\begin{equation}
\lambda ^{2}=\frac{g_{4}^{4}m_{s}^{12}V_{6}l^{4}}{(2\pi )^{14}}\text{ \ \ .}
\tag{3.12}
\end{equation}%
Substituting this expression into the resulting one on both sides of (2.2),
we receive the following equation in partial derivatives with respect to the
scale function $l(x)$: 
\begin{equation*}
\left[ \frac{(2\pi )^{7}}{m_{s}^{4}V_{6}g_{4}^{4}l^{2}}-M_{4}^{2}\right] R+%
\frac{(2\pi )^{7}}{m_{s}^{4}V_{6}g_{4}^{4}}\frac{(l-1)}{l^{2}}g^{AB}\left(
\Gamma _{AB}^{C}\Gamma _{CD}^{D}-\Gamma _{AC}^{D}\Gamma _{BD}^{C}\right) +
\end{equation*}%
\begin{equation}
+\frac{(2\pi )^{7}}{m_{s}^{4}V_{6}g_{4}^{4}}\frac{1}{l^{3}}\left[ \frac{%
\partial l}{\partial x^{C}}g^{AB}\Gamma _{AB}^{C}-\frac{\partial l}{\partial
x^{B}}g^{AB}\Gamma _{AC}^{C}\right] =0\text{ \ \ .}  \tag{3.13}
\end{equation}%
Note that for $l=1$ (the known gravitational theory) we obtain again
expression (2.2)\ for $M_{4}^{2}$. It turns out that for $l(x)$ we have both
the algebraic relation (3.6)\ and the above differential equation (3.13).

\section{(ANOTHER) \ ALGEBRAIC \ RELATION \ AND A \ QUASILINEAR \
DIFFERENTIAL \ EQUATION \ FROM \ THE \ \textquotedblright
COMPACTIFIED+RESCALED\textquotedblright\ \ LOW \ ENERGY \ TYPE \ I \ STRING
\ THEORY \ ACTION}

\bigskip This time we start from the action (2.8) and substitute the
expressions for the \textquotedblright unrescaled\textquotedblright\ scalar
quantities $F^{2}$ and $R$ 
\begin{equation}
F^{2}=\frac{1}{l^{2}}\widetilde{F}^{2}\text{ \ \ \ ; \ \ \ \ }R=\frac{1}{%
l^{4}}\widetilde{R}+\frac{(l^{2}-1)}{2l^{2}}g^{AC}g^{BD}(...)  \tag{4.1}
\end{equation}%
into the L. H. S. of (2.8).

Following the method, described in the previous section, we find for $%
\lambda $ 
\begin{equation}
\lambda =\frac{g_{4}m_{s}^{6}V_{6}}{(2\pi )^{7}l^{2}}\text{ \ \ \ ,} 
\tag{4.2}
\end{equation}%
which with respect to the function $l(x)$ can be considered as the
\textquotedblright dual\textquotedblright\ one, if compared with (3.5).
However, the obtained algebraic relation will be different from (3.6) 
\begin{equation}
\left[ \frac{(2\pi )^{7}}{V_{6}m_{s}^{4}g_{4}^{4}}-M_{(4)}^{2}\right] 
\widetilde{R}+\frac{(2\pi )^{7}l^{2}(l^{2}-1)}{2m_{s}^{4}V_{6}g_{4}^{4}}%
g^{AC}g^{BD}(....)=0\text{ \ \ .}  \tag{4.3}
\end{equation}%
If again expression (3.4) for $\widetilde{R}$ is used, the algebraic
relation (4.3) can be rewritten as 
\begin{equation}
\frac{1}{2}l^{2}(l^{2}-1)g^{AC}g^{BD}(....)=\frac{P^{2}R}{(P-NR)^{2}}+l^{4}R%
\text{ \ \ \ \ \ ,}  \tag{4.4}
\end{equation}%
where $P$ and $N$ denote the expressions 
\begin{equation}
P\equiv \frac{(2\pi )^{7}}{2m_{s}^{4}V_{6}g_{4}^{4}}g^{AC}g^{BD}(...)\text{
\ \ \ ; \ \ \ \ }N\equiv \frac{(2\pi )^{7}}{m_{s}^{4}V_{6}g_{4}^{4}}%
-M_{4}^{2}\text{ \ .}  \tag{4.5}
\end{equation}%
In the same way, starting from the second representation (3.11) of the
gravitational Lagrangian for $R$ in terms of $\ \widetilde{R}$ \ and again
making use of formulae (3.11) for $\widetilde{R}$, one can obtain the second
quasilinear equation in partial derivatives\textbf{\ } 
\begin{equation*}
\left[ \frac{(2\pi )^{7}}{m_{s}^{4}V_{6}g_{4}^{4}}l^{6}-M_{4}^{2}l^{4}\right]
R-\left[ \frac{(2\pi )^{7}l^{2}}{m_{s}^{4}V_{6}g_{4}^{4}}-M_{4}^{2}\right] 
\frac{l^{2}(l^{2}-1)}{2}g^{AC}g^{BD}(...)-
\end{equation*}%
\begin{equation*}
-\frac{(2\pi )^{7}l^{4}(l-1)}{m_{s}^{4}V_{6}g_{4}^{4}}g^{AB}\left( \Gamma
_{AB}^{C}\Gamma _{CD}^{D}-\Gamma _{AC}^{D}\Gamma _{BD}^{C}\right) -
\end{equation*}%
\begin{equation}
-\frac{(2\pi )^{7}l^{3}}{m_{s}^{4}V_{6}g_{4}^{4}}\left( \frac{\partial l}{%
\partial x^{C}}g^{AB}\Gamma _{AB}^{C}-\frac{\partial l}{\partial x^{B}}%
g^{AB}\Gamma _{AC}^{C}\right) =0\text{ \ \ \ }.  \tag{4.6}
\end{equation}%
Substituting the algebraic relation (4.4) into the second term of (4.6), the
differential equation is obtained in a simpler form 
\begin{equation*}
\frac{(2\pi )^{7}l^{3}}{m_{s}^{4}V_{6}g_{4}^{4}}\left( \frac{\partial l}{%
\partial x^{C}}g^{AB}\Gamma _{AB}^{C}-\frac{\partial l}{\partial x^{B}}%
g^{AB}\Gamma _{AC}^{C}\right) +\frac{RP^{2}}{l^{3}(P-NR)^{2}}\left[ \frac{%
(2\pi )^{7}l^{2}}{m_{s}^{4}V_{6}g_{4}^{4}}-M_{4}^{2}\right] +
\end{equation*}%
\begin{equation}
+\frac{(2\pi )^{7}l(l-1)}{m_{s}^{4}V_{6}g_{4}^{4}}g^{AB}\left( \Gamma
_{AB}^{C}\Gamma _{CD}^{D}-\Gamma _{AC}^{D}\Gamma _{BD}^{C}\right) =0\text{ \
\ \ \ . }  \tag{4.7}
\end{equation}%
This (second) differential equation evidently is different from the first
one (3.13) and in this aspect an interesting conclusion can be made. Suppose
that the two differential equations (3.13) and (4.7) simultaneously hold, 
\textit{which means that it does not matter in what sequence we perform
\textquotedblright rescaling + compactification\textquotedblright\ or
\textquotedblright compactification +rescaling\textquotedblright\ in the low
energy type I string theory action.} If the initial term with the
derivatives in (4.7) is expressed and substituted into the first
differential equation (3.13), then \textit{the square of the length scale
function }$l^{2}$ can be found as a solution of the following cubic
algebraic equation 
\begin{equation}
M_{4}^{2}l^{6}-\frac{(2\pi )^{7}}{m_{s}^{4}V_{6}g_{4}^{4}}l^{4}+\frac{(2\pi
)^{7}}{m_{s}^{4}V_{6}g_{4}^{4}}\frac{P^{2}}{(P-NR)^{2}}l^{2}-M_{4}^{2}\frac{%
P^{2}}{(P-NR)^{2}}=0\text{ \ \ .}  \tag{4.8}
\end{equation}

\textbf{\bigskip }Both the above mentioned approaches of \textit{%
\textquotedblright\ rescaling + compactification \textquotedblright }\ and 
\textit{\textquotedblright compactification + rescaling\textquotedblright }\
would be consistent in the case of a non - imaginary Lobachevsky space, if
the function $l(\mathbf{x})$ is a real one and not a complex one, i.e. the
roots of the above equation should not be imaginary functions and there
should be at least one root, which is a real function. This may lead
additionallly to some restrictions on the parameters in the initial string
action. Also, for $l=1$ the above equation can be written as 
\begin{equation}
\left[ M_{4}^{2}-\frac{(2\pi )^{7}}{m_{s}^{4}V_{6}g_{4}^{4}}\right] \left[ 
\frac{P^{2}}{(P-NR)^{2}}+1\right] =0\text{ \ \ .}  \tag{4.9}
\end{equation}

\bigskip There is no other relation from this equation besides the known one 
$M_{4}^{2}-\frac{(2\pi )^{7}}{m_{s}^{4}V_{6}g_{4}^{4}}=0$, since the
nominator of the second term can be written as 
\begin{equation}
2\left[ \left( \frac{P}{N}\right) ^{2}-\left( \frac{P}{N}\right) R+\frac{1}{2%
}R^{2}\right] =\frac{1}{2}\left[ \left( \frac{P}{N}-\frac{R}{2}\right) ^{2}+%
\frac{R^{2}}{4}\right] \text{ \ \ \ \ }  \tag{4.10}
\end{equation}%
and evidently this term is positive and different from zero.

However, if solutions of the two quasilinear differential equations are
found, then some new relations may be written. It should be kept in mind
that these solutions are found by means of the characteristic system of
equations, and the general solutions depend on the solutions of the
characteristic system.

\section{\protect\bigskip ALGEBRAIC \ INEQUALITIES \ FOR \ THE \ PARAMETERS
\ IN \ THE \ LOW - ENERGY \ TYPE \ I STRING \ THEORY \ ACTION \ \ }

\bigskip Taking into account expressions (4.5) for $P$ and $N$, eq.(4.8)
after dividing by $Q^{2}M_{4}^{2}m_{s}^{4}V_{6}g_{4}^{2}$ can be written in
the form of the following cubic algebraic equation with respect to the
variable $l_{1}=l^{2}$ 
\begin{equation}
l_{1}^{3}+a_{1}l_{1}^{2}+a_{2}l_{1}+a_{3}=0\text{ \ \ \ \ ,}  \tag{5.1}
\end{equation}%
where $Q$, $a_{1}$, $a_{2}$ and $a_{3}$ denote the expressions 
\begin{equation}
Q\equiv \frac{g^{AC}g_{BD}(2\pi )^{7}g_{4}^{4}(....)}{\left[
g^{AC}g^{BD}(...)(2\pi )^{7}g_{4}^{4}-2R\left( (2\pi
)^{7}-M_{4}^{2}V_{6}m_{s}^{4}g_{4}^{4}\right) \right] }\text{ \ \ \ ,} 
\tag{5.2}
\end{equation}%
\begin{equation}
a_{1}\equiv -\frac{(2\pi )^{7}}{M_{4}^{2}m_{s}^{4}V_{6}g_{4}^{2}}\text{ \ \
\ ; \ \ \ \ }a_{2}\equiv \frac{(2\pi )^{7}}{%
M_{4}^{2}m_{s}^{4}V_{6}g_{4}^{2}Q^{2}}\text{ \ \ ; \ \ }a_{3}\equiv -\frac{%
g_{4}^{2}}{Q^{2}}\text{ \ \ .}  \tag{5.3}
\end{equation}%
After the variable change $l_{1}=x-\frac{a_{1}}{3}$ equation (5.1) is
brought to the form 
\begin{equation}
x^{3}+ax+b=0\text{ \ \ , }  \tag{5.4}
\end{equation}%
where $a$ and $b$ are the expressions 
\begin{equation}
a\equiv a_{2}-\frac{a_{1}^{2}}{3}\text{ \ \ ; \ \ \ }b\equiv 2\frac{a_{1}^{3}%
}{27}-\frac{a_{1}a_{2}}{3}+a_{3}\text{ \ \ \ .}  \tag{5.5}
\end{equation}%
The roots of the cubic equation (5.4) are given by the formulae 
\begin{equation}
x=\sqrt[3]{p}-\frac{a}{3\sqrt[3]{p}}\text{ \ \ \ ,}  \tag{5.6}
\end{equation}%
where $p$ denotes the expression 
\begin{equation}
p\equiv -\frac{b}{2}\pm \sqrt{\frac{b^{2}}{4}+\frac{a^{3}}{27}}\text{ \ \ \ .%
}  \tag{5.7}
\end{equation}%
The roots of the cubic equation will not depend on the $+$ $\ $or $-$ sign
in front of the square in the above expression.

It may be noted that if the expression for $\frac{b^{2}}{4}+\frac{a^{3}}{27}$
is negative, then the corresponding roots $x_{1}$, $x_{2}$, $x_{3}$ and the
length function $l(x)$ will be imaginary. From a geometrical point of view,
this would be unacceptable, but with one exception - in the imaginary
Lobachevsky space [12], which is realized by all the straight lines outside
the absolute cone (on which the scalar product is zero, i.e. $[x,x]=0$), the
length may may take imaginary values in the interval $[0,\frac{\pi \text{ }i%
}{2k}]$ ($k$ is the Lobachevsky constant). Further we shall assume that $%
l(x) $ is a real function, but \textit{in principle it is interesting that
the sign of the inequalities, relating the parameters in the string action,
may change, if the spacetime is an imaginary Lobachevsky one.}

The expression for $\frac{b^{2}}{4}+\frac{a^{3}}{27}$ can be written as 
\begin{equation*}
\frac{b^{2}}{4}+\frac{a^{3}}{27}=\frac{1}{Q^{6}d^{6}}[\frac{1}{27}%
g_{4}^{6}d^{3}-\frac{1}{4.27}g_{4}^{8}d^{2}Q^{2}-\frac{2}{27^{2}}%
g_{4}^{12}Q^{6}+
\end{equation*}%
\begin{equation}
+\frac{1}{4}g_{4}^{2}Q^{2}-\frac{1}{6}g_{4}^{6}Q^{2}d^{4}+\frac{1}{27}%
g_{4}^{8}Q^{4}d^{3}]\text{ \ \ \ ,}  \tag{5.8}
\end{equation}%
where $d$ is the introduced notation for 
\begin{equation}
d\equiv \frac{M_{4}^{2}V_{6}m_{s}^{4}g_{4}^{4}}{\left( 2\pi \right) ^{7}}%
\text{ \ \ \ .}  \tag{5.9}
\end{equation}%
It is difficult to check when expression (5.8) will be non - negative, since 
$Q$ depends also on $d$ and a higher - degree polynomial with respect to $d$
will be obtained. However, it may be noted that since 
\begin{equation}
l^{2}=l_{1}=x-\frac{a_{1}}{3}>0\text{ \ \ }  \tag{5.10}
\end{equation}%
and since $a_{1}$ is a non - complex quantity, then all the roots $x_{1}$, $%
x_{2}$, $x_{3}$ are real. Therefore from the Wiet formulae 
\begin{equation}
-a=x_{1}+x_{2}+x_{3}>a_{1}  \tag{5.11}
\end{equation}%
and with account of the expressions for $a$ and $a_{1},$ an equality can be
obtained with respect to $d$ and the electromagnetic constang $g_{4}$ in the
type I low energy string theory acton 
\begin{equation}
\frac{1}{3}g_{4}^{2}>W(W+2)d^{4}-2W(W+1)d^{3}+W^{2}d^{2}\text{ \ \ \ \ \ ,} 
\tag{5.12}
\end{equation}%
where $W$ is the notation for 
\begin{equation}
W\equiv \frac{2R}{g^{AB}g^{CD}(...)g_{4}^{4}}\text{ \ \ \ .}  \tag{5.13}
\end{equation}%
The last (third) inequality with the parameters of the low - energy type I
string theory action can be obtained from the restriction (5.10) $x>\frac{%
a_{1}}{3}$ for the roots of the cubic equation and expression (5.6) for $x$ 
\begin{equation}
\frac{3\sqrt[3]{p^{2}}-a}{3\sqrt[3]{p}}>\frac{a_{1}}{3}\text{ \ \ \ .} 
\tag{5.14}
\end{equation}%
Denoting 
\begin{equation}
q_{1}=\sqrt[3]{p^{2}}\text{ \ \ ,}  \tag{5.15}
\end{equation}%
the above inequality can be rewritten as 
\begin{equation}
9q_{1}^{2}-(a_{1}^{2}+6a)q_{1}+a^{2}>0\text{ \ \ \ \ \ \ ,}  \tag{5.16}
\end{equation}%
which is satisfied for 
\begin{equation}
p^{2}=\frac{b^{2}}{2}+\frac{a^{3}}{27}-b\sqrt{\frac{b^{2}}{2}+\frac{a^{3}}{27%
}}>\left[ \frac{a_{1}+6a}{18}+\frac{a_{1}}{18}\sqrt{a_{1}^{2}+12a}\right]
^{3}  \tag{5.17}
\end{equation}%
or for 
\begin{equation}
p^{2}=\frac{b^{2}}{2}+\frac{a^{3}}{27}-b\sqrt{\frac{b^{2}}{2}+\frac{a^{3}}{27%
}}<\left[ \frac{a_{1}+6a}{18}-\frac{a_{1}}{18}\sqrt{a_{1}^{2}+12a}\right]
^{3}\text{ \ \ .}  \tag{5.18}
\end{equation}%
\textit{The last two inequalities are the new inequalities between the
parameters in the low-energy type I string theory action, which cannot be
obtained in the framework of the usual gravity theory.} Note that in these
inequalities the length function $l(x)$ no longer appears, so the obtained
result should be valid also for the case of standard gravitational theory.

\section{\protect\bigskip SOLUTIONS \ OF \ THE \ FIRST \ QUASILINEAR \
DIFFERENTIAL \ EQUATION \ (3.13) \ FOR \ THE \ CASE \ OF \ A \ FLAT \ $4D$ \
MINKOWSKI \ METRIC, \ EMBEDDED \ IN \ A \ FIVE - DIMENSIONAL SPACE (OF \
CONSTANT \ NEGATIVE \ OR \ NON - CONSTANT \ CURVATURE)\ }

\bigskip The purpose will be to show that the differential equation in
partial derivatives (3.13) will be solvable for the previously considered
case of the metric (2.4), written now as 
\begin{equation}
ds^{2}=e^{-2k\epsilon y}\eta _{\mu \nu }dx^{\mu }dx^{\nu }+h(y)dy^{2}\text{
\ \ \ \ \ \ ,}  \tag{6.1}
\end{equation}%
$\eta _{\mu \nu }=(+,-,-,-)$ with $h(y)=1$ and $\epsilon =\pm 1$. Moreover,
the equation will turn out to be solvable also for the case of a power -
like dependence of the scale factor $h(y)=\gamma y^{n}$ ($\gamma =const$),
when the five - dimensional scalar curvature is no longer a constant one. In
principle, it is necessary to know for what kind of metrics quasilinear
differential equations of the type (3.13) admit exact analytical solutions.
It may be shown that for more complicated metrics (for example, when the
embedded four - dimensional spacetime is a Schwarzschild Black hole with an
warp factor) such analytical solutions cannot be found. This will be shown
elsewhere.

As usual, the Greek indices $\mu ,\nu ,\alpha ,\beta $ will run from $1$ to $%
4$ and the extra - dimensional metric tensor component is $h(y)\equiv g_{55}$%
. The big letters $A,B,C...$ will denote the coordinates of the five -
dimensional spacetime.

The corresponding affine connection components are 
\begin{equation}
\Gamma _{\mu \nu }^{\alpha }=\Gamma _{5\nu }^{\alpha }=0\text{ \ \ \ ; \ \ }%
\Gamma _{\mu \nu }^{5}=\frac{k\varepsilon }{h}\eta _{\mu \nu
}e^{-2k\varepsilon y}\text{ \ \ \ ,}  \tag{6.2}
\end{equation}%
\begin{equation}
\Gamma _{\mu 5}^{5}=0\text{ \ ;\ \ }\Gamma _{55}^{5}=\frac{1}{2}\frac{%
h^{^{\prime }}(y)}{h(y)}\text{ \ \ ; \ \ }\Gamma _{55}^{\alpha }=\frac{1}{2}%
e^{2k\varepsilon y}\eta _{\alpha \alpha }h^{^{\prime }}(y)\text{ \ \ \ .} 
\tag{6.3}
\end{equation}%
The expressions for the scalar curvature $R$ and for $g^{AB}(\Gamma
_{AB,C}^{C}-\Gamma _{AC,B}^{C})$ are 
\begin{equation}
R=-\frac{8k^{2}}{h}-4k\epsilon \frac{h^{^{\prime }}}{h^{2}}\text{ }%
=g^{AB}(\Gamma _{AB,C}^{C}-\Gamma _{AC,B}^{C})\text{\ \ ,}  \tag{6.4}
\end{equation}%
from where, taking the difference of the two expressions, it is found that $%
g^{AB}(\Gamma _{AB}^{C}\Gamma _{CD}^{D}-\Gamma _{AC}^{D}\Gamma _{BD}^{C})=0$
and the differential equation (3.13) is written as 
\begin{equation*}
\frac{(2\pi )^{7}}{m_{s}^{4}V_{5}g_{5}^{5}}\frac{e^{2k\epsilon y}h^{^{\prime
}}}{2h}\left[ \frac{\partial l}{\partial x^{1}}-\frac{\partial l}{\partial
x^{2}}-\frac{\partial l}{\partial x^{3}}-\frac{\partial l}{\partial x^{4}}%
\right] +
\end{equation*}%
\begin{equation}
+\frac{(2\pi )^{7}4k\epsilon }{m_{s}^{4}V_{5}g_{5}^{4}h}\frac{\partial l}{%
\partial y}=Cl-Dl^{3}\text{ \ ,}  \tag{6.5}
\end{equation}%
where $C$ and $D$ denote the expressions 
\begin{equation}
C\equiv \frac{(2\pi )^{7}4k(2kh+\epsilon h^{^{\prime }})}{%
m_{s}^{4}V_{5}g_{5}^{4}h^{2}}\text{ \ \ ; \ \ }D\equiv \frac{%
M_{5}^{2}4k(2kh+\epsilon h^{^{\prime }})}{h^{2}}\text{ \ \ \ .}  \tag{6.6}
\end{equation}%
The characteristic system of equations for the equation (6.5) is 
\begin{equation}
\frac{dl}{Cl-Dl^{3}}=\frac{\varepsilon m_{s}^{4}V_{5}g_{5}^{4}h}{(2\pi
)^{7}4k}dy=d\sigma \text{ \ \ \ ,}  \tag{6.7}
\end{equation}%
\begin{equation}
\frac{2m_{s}^{4}V_{5}g_{5}^{4}e^{-2k\varepsilon y}}{(2\pi )^{7}(\ln
h)^{^{\prime }}}dx^{1}=-\frac{2m_{s}^{4}V_{5}g_{5}^{4}e^{-2k\varepsilon y}}{%
(2\pi )^{7}(\ln h)^{^{\prime }}}dx^{i}=d\sigma \text{ \ \ ,}  \tag{6.8}
\end{equation}%
where the indice $i=2,3,4$ and $\sigma $ is some parameter. The solution of
the first characteristic equation for the $y$ and $l$ variables is 
\begin{equation}
\frac{\varepsilon _{1}l}{\left[ \varepsilon _{2}(l^{2}-\alpha _{1}^{2})%
\right] ^{\frac{1}{2}}}=C_{1}(x_{1},x_{i})e^{2k\varepsilon _{3}y}h\text{ \ \
\ \ ,}  \tag{6.9}
\end{equation}%
where 
\begin{equation}
\alpha _{1}=\sqrt{\frac{C}{D}}\text{ \ \ \ \ }  \tag{6.10}
\end{equation}%
and $\varepsilon _{1}$, $\varepsilon _{2}$, $\varepsilon _{3}$ take values $%
\pm 1$ independently one from another.

In order to find the function $C_{1}(x_{1},x_{i})$, the obtained solution
(6.9) should be differentiated by $x_{1}$ and the characteristic equations
for $\frac{\partial l}{\partial y}$ and $\frac{\partial y}{\partial x_{1}}$
have to be used. As a result, the function $C_{1}(x_{1},x_{i})$ is found as
a solution of the following simple differential equation 
\begin{equation}
M=S\frac{\partial C_{1}(x_{1},x_{i})}{\partial x_{1}}+TC_{1}(x_{1},x_{i})%
\text{ \ \ \ \ ,}  \tag{6.11}
\end{equation}%
where the functions $M$, $S$ and $T$ are defined as follows 
\begin{equation}
M\equiv -\frac{\varepsilon _{1}(2l^{2}-\alpha
_{1}^{2})l8kM_{5}^{2}(2kh+\varepsilon _{3}h^{^{\prime
}})m_{s}^{4}V_{5}g_{5}^{4}}{\left[ \varepsilon _{2}(l^{2}-\alpha _{1}^{2})%
\right] ^{\frac{1}{2}}(2\pi )^{7}hh^{^{\prime }}e^{2k\varepsilon _{3}y}}%
\text{ \ for \ }\varepsilon _{1}\varepsilon _{2}=-1\text{ \ ,}  \tag{6.12}
\end{equation}%
\begin{equation}
M\equiv \frac{\varepsilon _{1}l\alpha _{1}^{2}8kM_{5}^{2}(2kh+\varepsilon
_{3}h^{^{\prime }})m_{s}^{4}V_{5}g_{5}^{4}}{\left[ \varepsilon
_{2}(l^{2}-\alpha _{1}^{2})\right] ^{\frac{1}{2}}(2\pi )^{7}hh^{^{\prime
}}e^{2k\varepsilon _{3}y}}\text{ \ for \ \ }\varepsilon _{1}\varepsilon
_{2}=+1\text{ \ \ \ \ ,}  \tag{6.13}
\end{equation}%
\begin{equation}
S\equiv he^{2k\varepsilon _{3}y}\text{ \ \ ; \ \ \ }T\equiv \frac{%
8k(2k+\varepsilon _{3}h^{^{\prime }})}{h^{^{\prime }}}\text{ \ \ \ \ .} 
\tag{6.14}
\end{equation}%
The solution of the differential equation (6.11) can be written as 
\begin{equation}
C_{1}(x_{1},x_{i})=\frac{M}{T}-\varepsilon _{4}\frac{C_{2}(x_{i})}{T}%
e^{-\int \frac{T}{S}dx_{1}}\text{ \ \ .}  \tag{6.15}
\end{equation}%
Again, the obtained solution (6. 9) can be differentiated with respect to $%
x_{i}$ and taking into account from the characteristic equations that 
\begin{equation}
\frac{\partial l}{\partial x_{i}}=-\frac{\partial l}{\partial x_{1}}\text{ \
\ \ ; \ \ \ }\frac{\partial y}{\partial x_{i}}=-\frac{\partial y}{\partial
x_{1}}\text{ \ \ \ ,}  \tag{6.16}
\end{equation}%
the solution of the corresponding equation (6.11) with $\widetilde{M}=-M$, $%
\widetilde{T}=-T$, $\widetilde{S}=S$ can be represented as 
\begin{equation}
C_{1}(x_{1},x_{i})=\frac{M}{T}+\varepsilon _{4}\frac{C_{3}(x_{1})}{T}e^{\int 
\frac{T}{S}dx_{i}}\text{ \ \ .}  \tag{6.17}
\end{equation}%
Substracting the two relations (6.15) and (6.17), the following relation
between the functions $C_{2}(x_{i})$ and $C_{3}(x_{1})$ can be derived 
\begin{equation}
C_{2}(x_{i})=e^{\int \frac{T}{S}dx_{1}}C_{3}(x_{1})e^{\int \frac{T}{S}dx_{i}}%
\text{ \ \ \ \ .}  \tag{6.18}
\end{equation}%
Now let us differentiate relation (6.17) with respect to $x_{1}$ and make
use of (6.18) and its derivative also with respect to $x_{1}$. Then the
following differential equation is derived with respect to the function $%
C_{3}(x_{1})$ 
\begin{equation}
\frac{\partial C_{3}(x_{1})}{\partial x_{1}}-C_{3}(x_{1})\frac{T}{S}=0\text{
\ \ \ \ ,}  \tag{6.19}
\end{equation}%
from where 
\begin{equation}
C_{3}(x_{1})=const.\text{ }e^{^{\int \frac{T}{S}dx_{1}}}\text{ \ \ .} 
\tag{6.20}
\end{equation}%
Therefore from (6.17) 
\begin{equation}
C_{1}(x_{1},x_{i})=\frac{M+\varepsilon _{4}}{T}  \tag{6.21}
\end{equation}%
and substituting into (6.9), a final expression for $l$ can be found (for
the case $\varepsilon _{1}\varepsilon _{2}=+1$) 
\begin{equation}
l^{2}=\frac{\varepsilon _{2}\alpha _{1}^{2}e^{4\varepsilon _{3}ky}h^{2}}{%
\varepsilon _{2}h^{2}e^{4\varepsilon _{3}ky}-\frac{T^{2}}{(M+\varepsilon
_{4})^{2}}}\text{ \ \ \ .}  \tag{6.22}
\end{equation}%
The solution for the other case $\varepsilon _{1}\varepsilon _{2}=-1$ can be
found analogously.

The general solution of the quasilinear differential equation in partial
derivativatives will be not only expression (6.22), but also any function $V$%
, depending on the first integrals $K_{1},K_{2},K_{3}$,....,$K_{6}$ of the
characteristic system of equations [13] 
\begin{equation}
V=V(K_{1},K_{2},K_{3},K_{4},K_{5},K_{6})\text{ \ \ \ \ .}  \tag{6.23}
\end{equation}%
Now let us find a solution of the characteristic equation for the $x_{i}$
and $y$ variables for the case of the function $h(y)=\gamma y^{n}$. The
equation can be written as 
\begin{equation}
e^{2k\varepsilon _{4}y}\text{ }y^{n-1}=-\varepsilon _{4}\frac{8k}{n\gamma }%
dx^{i}\text{ \ \ ,}  \tag{6.24}
\end{equation}%
from where $x^{i}$ can be expressed as 
\begin{equation}
x^{i}=-\varepsilon _{4}\frac{n\gamma }{8k}I_{n-1}(k,y)\text{ \ }  \tag{6.25}
\end{equation}%
and $I_{n-1}(k,y)$ denotes the integral 
\begin{equation}
I_{n-1}(k,y)=\int e^{2k\varepsilon _{4}y}y^{n-1}dy\text{ \ \ .}  \tag{6.26}
\end{equation}%
This integral can be exactly calculated and the result can be found in the
monograph by Timofeev [14] 
\begin{equation}
I_{n-1}=\frac{e^{2k\varepsilon _{4}\text{ }y}}{(2k\varepsilon _{4})^{n}\text{
}}\sum\limits_{p=0}^{n-1}(-1)^{p}\left( 
\begin{array}{c}
n-1 \\ 
p%
\end{array}%
\right) p!(2k\varepsilon _{4})^{n-1-p}y^{n-1-p}\text{ \ \ \ \ .}  \tag{6.27 }
\end{equation}%
Note that because of the complicated expression for the integral $%
I_{n-1}(k,y)$, $y$ cannot be expressed as a function of the $x_{i}$ and the $%
x_{1}$ coordinate. Also, the solvability of the quasilinear differential
equation is determined mostly by the presence of the embedded flat $4D$
Minkowski spacetime. Therefore, it may be expected that there might be
another functions $h(y)$, for which exact analytical solution may be found.

\section{\protect\bigskip SOLUTIONS \ OF \ THE \ SECOND \ QUASILINEAR \
DIFFERENTIAL \ EQUATION \ (4.6) \ FOR \ THE \ CASE \ OF \ A \ FLAT \ $4D$ \
MINKOWSKI \ METRIC, \ EMBEDDED \ IN \ A \ FIVE - DIMENSIONAL SPACE \ }

\bigskip The same approach, developed in the previous section, shall be
applied with respect to the second quasilinear differential equation in
partial derivatives (4.6). The aim will be to show that the analytical \
solution will be different, compared to the first one for the differential
equation (3.13).

The differential equation (4.6) for the case of the metric (6.1) can be
written as 
\begin{equation}
-D\frac{\partial l}{\partial x^{1}}+D\left[ \frac{\partial l}{\partial x^{2}}%
+\frac{\partial l}{\partial x^{3}}+\frac{\partial l}{\partial x^{4}}\right]
+E\frac{\partial l}{\partial y}+Al^{4}+Bl^{2}+C=0\text{ \ \ \ ,}  \tag{7.1}
\end{equation}%
where $A,B,C,D,E$ denote the expressions 
\begin{equation}
A\equiv \frac{(2\pi )^{7}4k(2kh-\varepsilon h^{^{\prime }})}{%
h^{2}m_{s}^{4}V_{5}g_{s}^{4}}\text{ \ \ ; \ \ }C\equiv \frac{16k^{2}M_{5}^{2}%
}{h}\text{ \ \ \ \ ,}  \tag{7.2}
\end{equation}%
\begin{equation}
B\equiv \frac{16k^{2}(2\pi )^{7}}{hm_{s}^{4}g_{5}^{4}V_{5}}+\frac{4kM_{5}^{2}%
}{h^{2}}(\varepsilon h^{^{\prime }}-2kh)\text{ \ \ \ \ ,}  \tag{7.3}
\end{equation}%
\begin{equation}
D\equiv -\frac{(2\pi )^{7}le^{2k\varepsilon y}h^{^{\prime }}}{%
m_{s}^{4}V_{5}g_{5}^{4}2h}\text{ \ \ \ ; \ \ \ }E\equiv -\frac{4k\varepsilon
(2\pi )^{7}l}{m_{s}^{4}V_{5}g_{5}^{4}h}\text{ \ \ \ .}  \tag{7.4}
\end{equation}%
The characteristic system of equations is 
\begin{equation}
\frac{dx_{1}}{D}=-\frac{dx_{i}}{D}=-\frac{dy}{E}=\frac{dl}{Al^{4}+Bl^{2}+C}%
\text{ \ \ \ .}  \tag{7.5}
\end{equation}%
The characteristic equation for the $y$ and $l$ variables can be written as 
\begin{equation}
d\left[ \ln \left( \varepsilon _{1}\left( \frac{s-G}{s+G}\right) \right) %
\right] =2AG\frac{\varepsilon _{4}hm_{s}^{4}V_{5}g_{5}^{4}}{4k(2\pi )^{7}}dy%
\text{ \ \ \ ,}  \tag{7.6}
\end{equation}%
where 
\begin{equation}
G\equiv \frac{B^{2}}{4A^{2}}-\frac{C}{A}=\frac{M_{5}\sqrt{(F-\frac{k}{2}%
)^{2}+k^{2}\left( 2(2\pi )^{7}-\frac{1}{4}\right) }}{\sqrt{\pi }(2\pi
)^{3}\mid 2k-\varepsilon _{4}\frac{h^{^{\prime }}}{h}\mid }\text{ \ \ } 
\tag{7.7}
\end{equation}%
is a non - negative function and 
\begin{equation}
s\equiv l^{2}+\frac{B}{2A}\text{ \ \ ; \ \ \ }F\equiv
m_{s}^{4}V_{5}g_{5}^{4}(2k-\varepsilon _{4}\frac{h^{^{\prime }}}{h})\text{ \
\ \ \ .}  \tag{7.8}
\end{equation}%
The integration of the characteristic equation (7.6) results in the
expression 
\begin{equation}
l^{2}=-\frac{B}{2A}+G\frac{(1+\varepsilon _{1}D_{1}(x_{1},x_{i})e^{Z\text{ }%
\widetilde{J}(y)})}{(1-\varepsilon _{1}D_{1}(x_{1},x_{i})e^{Z\text{ }%
\widetilde{J}(y)}}\text{ \ \ \ \ \ ,}  \tag{7.9}
\end{equation}%
where $Z$ is given by 
\begin{equation}
Z:=\frac{2M_{5}\varepsilon _{4}m_{s}^{2}g_{5}^{2}\sqrt{V_{5}}}{(2\pi )^{3}} 
\tag{7.10}
\end{equation}%
and $\widetilde{J}(y)$ is the integral 
\begin{equation}
\widetilde{J}(y):=\int \frac{\sqrt{K_{1}y^{2}+K_{2}y+1}}{y}dy\text{ \ \ \ .}
\tag{7.11}
\end{equation}%
The functions $K_{1}$ and $K_{2}$ are the following 
\begin{equation}
K_{1}\equiv \frac{k^{2}\left[ 2(2\pi )^{7}-\frac{1}{4}\right] }{%
m_{s}^{8}V_{5}^{2}g_{5}^{8}}+k^{2}\left( 2-\frac{1}{2m_{s}^{4}V_{5}g_{5}^{4}}%
\right) ^{2}\text{ \ \ ,}  \tag{7.12}
\end{equation}%
\begin{equation}
K_{2}\equiv -2\varepsilon _{4}\left( 2k-\frac{k}{2m_{s}^{4}V_{5}g_{5}^{4}}%
\right) \text{ \ \ \ .}  \tag{7.13}
\end{equation}%
It is important to note that the free term under the square in the integral $%
\widetilde{J}(y)$ (7.11) is positive (it is $+1$) and the function $K_{1}$
is also positive. The analytical solution of integrals of the type (7.11)
depends on the sign of the function $K_{1}$ and of the free term, which in
the present case are both positive. The explicite solution can be found in
the book of Timofeev [14]: 
\begin{equation*}
\widetilde{J}(y)\equiv \sqrt{K_{1}y^{2}+K_{2}y+1}+\frac{K_{2}}{2\sqrt{K_{1}}}%
\ln (K_{1}y+\frac{K_{2}}{2}+
\end{equation*}%
\begin{equation}
+\sqrt{K_{1}}\sqrt{K_{1}y^{2}+K_{2}y+1})-\ln \left( \frac{1+\frac{K_{2}}{2}y+%
\sqrt{K_{1}y^{2}+K_{2}y+1}}{y}\right) \text{ \ \ \ \ \ .}  \tag{7.14}
\end{equation}%
Now it remains to determine the function $D_{1}(x_{1},x_{i})$ in expression
(7.9) for $l^{2}$. For the purpose, let us denote the under - integral
expression in (7.11) by $J(y)$, differentiate both sides of (7.9) by $x_{1}$
and take into account the expressions for $\frac{\partial l}{\partial x_{1}}$
and $\frac{\partial y}{\partial x_{1}}$ from the characteristic system of
equations. After rearranging the terms and denoting by $V$%
\begin{equation}
V\equiv \varepsilon _{1}\frac{-\frac{2l(Al^{4}+Bl^{2}+C)}{E}+\frac{\partial 
}{\partial y}\left( \frac{B}{2A}\right) -\frac{\partial Q}{\partial y}\frac{%
(l^{2}+\frac{B}{2A})}{G}}{G+l^{2}+\frac{B}{2A}}\text{ \ \ \ \ ,}  \tag{7.15}
\end{equation}%
the following differential equation can be obtained for the function $%
D_{1}(y)$ 
\begin{equation}
\frac{\partial D_{1}}{\partial y}+D_{1}(ZJ(y)+V)-Ve^{-Z\widetilde{J}(y)}=0%
\text{ \ \ \ .}  \tag{7.16}
\end{equation}%
It may seem strange at first glance that the function $D_{1}$ depends on the 
$y$ coordinate, while in (7.9) it was assumed that $D_{1}=D_{1}(x_{1},x_{i})$%
. In fact, from the characteristic equations (7.5) for the $y$ and $x_{1}$
variables it follows 
\begin{equation}
\frac{dy}{e^{2k\varepsilon _{4}y}h^{^{\prime }}}=-\frac{\varepsilon _{4}}{8k}%
dx_{1}\text{ \ \ .}  \tag{7.17}
\end{equation}%
For the concrete expression for the function $h(y)=\gamma y^{n}$, the
coordinates $x_{1}$ and $x_{i}$ can be expressed as 
\begin{equation}
x_{1}=-\varepsilon _{4}\frac{8k}{n\gamma }I(-k,1-n)+const.\text{ \ ; \ }%
x_{i}=-x_{1}  \tag{7.18}
\end{equation}%
and therefore it is reasonable to consider that $%
D_{1}=D_{1}(x_{1}(y),x_{i}(y))=D_{1}(y)$. Note however that due to the
complicated structure of the integral $I(-k,1-n)$, it is impossible to
express $y$ as a function of $x_{1}$ (or $x_{i}$).

As in the previous case, the general solution of the equation depends on all
the first integrals of the characteristic system of equations.

\section{\protect\bigskip LENGTH \ FUNCTION \ $l(x)$ \ FROM \ THE \
CONSTANCY \ OF \ THE \ SCALAR \ CURVATURE $\ R$ \ UNDER \ \textquotedblright
RESCALINGS\textquotedblright\ \ OF \ THE \ CONTRAVARIANT \ METRIC \ TENSOR \
FOR \ THE \ CASE \ OF \ A \ FLAT \ $4D$ \ MINKOWSKI \ METRIC, \ EMBEDDED \
IN A \ $5D$ \ SPACETIME. \ }

\bigskip Now we shall find solutions of the corresponding differential
equation in partial derivatives, when the second representation of the
scalar curvature $\widetilde{R}$ \ (3.11) is equal to the initial scalar
curvature $R$ (i. e. $\widetilde{R}=R$). The obtained differential equation
under this identification is 
\begin{equation*}
l^{3}\left[ R-\frac{1}{2}g^{AC}g^{BD}(...)\right] +l^{2}\left[
-R+g^{AB}(\Gamma _{AB,C}^{C}-\Gamma _{AC,B}^{C})\right] +
\end{equation*}%
\begin{equation*}
+l\left[ \frac{1}{2}g^{AC}g^{BD}(...)-g^{AB}(\Gamma _{AB,C}^{C}-\Gamma
_{AC,B}^{C})\right] +
\end{equation*}%
\begin{equation}
+\frac{\partial l}{\partial x^{B}}g^{AB}\Gamma _{AC}^{C}-\frac{\partial l}{%
\partial x^{C}}g^{AB}\Gamma _{AB}^{C}=0\text{ \ \ \ \ .}  \tag{8.1}
\end{equation}%
The expression in the small brackets is the same as in (3.4), i.e. $%
(...)\equiv (g_{AD,BC}+g_{BC,AD}-g_{AC,BD}-g_{BD,AC})$. The equation (8.1)
for the case of the metric (6.1) with the affine connection components (6.2)
- (6.3) acquires the form 
\begin{equation*}
\varepsilon \frac{\partial l}{\partial y}+\frac{h^{^{\prime }}}{8k}%
e^{2k\varepsilon y}\left( \frac{\partial l}{\partial x_{1}}-\frac{\partial l%
}{\partial x_{2}}-\frac{\partial l}{\partial x_{3}}-\frac{\partial l}{%
\partial x_{4}}\right) =
\end{equation*}%
\begin{equation}
=(2k-\varepsilon \frac{h^{^{\prime }}}{h})(l^{3}-l)\text{ \ \ .}  \tag{8.2}
\end{equation}%
The characteristic system of equations is 
\begin{equation}
\frac{dl}{(2k-\varepsilon \frac{h^{^{\prime }}}{h})l(l^{2}-1)}=\varepsilon
dy=  \tag{8.3}
\end{equation}%
\begin{equation}
=\frac{dx_{1}}{h^{^{\prime }}}8ke^{-2k\varepsilon y}=-\frac{dx_{i}}{%
h^{^{\prime }}}8ke^{-2k\varepsilon y}\text{ \ \ .}  \tag{8.4}
\end{equation}%
The solutions of the characteristic system for the $x_{1}$ and $y$ variables
are correspondingly 
\begin{equation}
x_{1}=C_{4}(x_{i},l)+\varepsilon _{2}\frac{e^{2k\varepsilon _{1}y}h}{8k}-%
\frac{1}{4}\int he^{2k\varepsilon _{1}y}dy\text{ \ \ ,}  \tag{8.5}
\end{equation}%
\begin{equation}
l^{2}=\frac{h^{2}}{h^{2}-D_{2}(x_{1},x_{i})e^{4k\varepsilon _{1}y}}\text{ \
\ \ .}  \tag{8.6}
\end{equation}%
Unfortunately, (8.6) cannot be considered as an expression for $l$, since
from (8.5) it is obvious that $D_{2}(x_{1},x_{i})$ also depends on the
function $l$. Also, it should be understood that $D_{2}$ depends on all the
variables $x_{i}$, $i=2,3,4$.

Let us differentiate both sides of (8.6) by $x_{1}$ 
\begin{equation*}
2l\frac{\partial l}{\partial x_{1}}=\frac{2lh^{^{\prime }}}{h}\frac{\partial
y}{\partial x_{1}}-\frac{l^{4}}{h^{2}}(2hh^{^{\prime }}\frac{\partial y}{%
\partial x_{1}}-
\end{equation*}%
\begin{equation}
-2D_{2}\frac{\partial D_{2}}{\partial x_{1}}e^{4k\varepsilon
_{1}y}-D_{2}4k\varepsilon _{1}e^{4k\varepsilon _{1}y}\frac{\partial y}{%
\partial x_{1}})\text{ \ \ .}  \tag{8.7}
\end{equation}%
If the same operation is applied also with respect to the $x_{i}$ coordinate
and the derived equation is summed up with (8.7) with account also of $\frac{%
\partial l}{\partial x_{1}}=-\frac{\partial l}{\partial x_{i}}$, $\frac{%
\partial y}{\partial x_{1}}=-\frac{\partial y}{\partial x_{i}}$, then it can
be obtained 
\begin{equation}
\frac{2l^{4}}{h^{2}}e^{4k\varepsilon _{1}y}D_{2}\left( \frac{\partial D_{2}}{%
\partial x_{1}}+\frac{\partial D_{2}}{\partial x_{i}}\right) =0\text{ \ \ \
\ .}  \tag{8.8}
\end{equation}%
The equation is satisfied also for $D_{2}\equiv 0$, which evidently
corresponds to the standard case $l=1$ in gravity theory. The other case,
when the equation is fulfilled, is $\frac{\partial D_{2}}{\partial x_{1}}=-%
\frac{\partial D_{2}}{\partial x_{i}}$.

Now let us rewrite equation (8.7) with account of the expressions for $\frac{%
\partial l}{\partial x_{1}}$ and $\frac{\partial y}{\partial x_{1}}$ from
the characteristic system of equations (8.3 - 8.4). Then the following
qusilinear differential equation with respect to the function $D_{2}^{2}$ is
derived 
\begin{equation*}
\frac{\partial D_{2}^{2}}{\partial y}-4k\varepsilon _{1}(1+\frac{4}{h}%
)D_{2}^{2}-
\end{equation*}%
\begin{equation}
-2\varepsilon _{1}\varepsilon _{5}\frac{h^{^{\prime }}}{h}(2k-\varepsilon
_{1}\frac{h^{^{\prime }}}{h})e^{-4k\varepsilon
_{1}y}(h^{2}-D_{2}^{2}e^{4k\varepsilon _{1}y})^{\frac{3}{2}}=0\text{ \ \ \ \
.}  \tag{8.9}
\end{equation}%
After finding the function $D_{2}=D_{2}(y)$ as a solution of this equation,
from (8.6) $l^{2}$ can also be found as a function of the extra - coordinate 
$y$, i.e. $l=l(y)$.Then after differentiating the solution for $x_{1}$ (8.5)
by $y$, the function $C_{4}(x_{i},l)$ can be derived as a solution of the
following differential equation 
\begin{equation}
\frac{\partial C_{4}(x_{i},l)}{\partial y}=F_{1}\text{ \ ,}  \tag{8.10}
\end{equation}%
where the function $F_{1}$ is determined as 
\begin{equation}
F_{1}\equiv e^{2k\varepsilon _{1}y}[\frac{\varepsilon
_{2}l(l^{2}-1)h^{^{\prime }}}{8k}-\frac{\varepsilon _{1}\varepsilon _{2}h}{4}%
-\frac{\varepsilon _{2}h^{^{\prime }}}{8k}+\frac{h}{4}]  \tag{8.11}
\end{equation}%
and $l$ has to be substituted with expression (8.6), in which $D_{2}$ is
determined as a solution of the differential equation (8.9). The
representation in the form (8.10) is particularly convenient for the case $%
h(y)=\gamma y^{n}$, when the difficulty will be only in calculating the
integral along $y$ in the first term of (8.11). The advantage of the
representation (8.10) will become evident if we differentiate the solution
(8.5) by $l$, obtaining thus the differential equation 
\begin{equation}
E_{1}(y)=\frac{\partial C_{4}(x_{i},l)}{\partial l}+\frac{E_{2}(y)}{%
l(l^{2}-1)}\text{ \ \ ,}  \tag{8.12}
\end{equation}%
where 
\begin{equation}
E_{1}(y)\equiv \frac{hh^{^{\prime }}e^{2k\varepsilon _{1}y}}{%
8k(2kh-\varepsilon _{2}h^{^{\prime }})}\text{ \ \ \ ,}  \tag{8.13}
\end{equation}%
\begin{equation}
E_{2}(y)\equiv \frac{\varepsilon _{2}he^{2k\varepsilon _{1}y}}{%
(2kh-\varepsilon _{2}h^{^{\prime }})}\left[ \left( \varepsilon
_{1}\varepsilon _{2}-1\right) +\frac{\varepsilon _{2}h^{^{\prime }}}{8k}%
\right] \text{ \ \ .}  \tag{8.14}
\end{equation}%
Now it is important to stress that the second representation (8.12) is
inconvenient to use for the case $h(y)=\gamma y^{n}$. The reason is that the
integration is along the $l$ coordinate, which means that the $y$ coordinate
in $E_{1}(y)$ and $E_{2}(y)$ has to be expressed from expression (8.6) as a
function of $l$. However, in view of the extremely complicated expression,
this is not possible. Instead, the differential equation (8.12) will be very
helpful for the case $h(y)\equiv 1$, which is frequently encountered in most
of the papers on theories with extra dimensions. Indeed, then the nonlinear
differential equation (8.9) is of a particularly simple form: 
\begin{equation}
\frac{\partial D_{2}^{2}}{\partial y}-20k\varepsilon D_{2}^{2}=0\text{ \ \ ,}
\tag{8.15}
\end{equation}%
from where with the help of (8.6) 
\begin{equation}
l^{2}=\frac{1}{1-const.e^{24k\varepsilon _{1}y}}\text{ \ \ \ .}  \tag{8.16}
\end{equation}%
Note one interesting property of the obtained solution, already mentioned in
the Introduction - when $\varepsilon _{1}=-1$\ and $y$\ tends to infinity,
the known case in gravity theory $l^{2}=1$\ is recovered.

Now $y$ can be expressed easily and the resulting differential equation
(8.12) with $E_{1}(y)=0$ can be rewritten as 
\begin{equation}
\frac{\partial C_{4}(x_{i},l)}{\partial l}+\frac{\varepsilon
_{2}(\varepsilon _{1}\varepsilon _{2}-1)e^{\frac{1}{12}}}{const.8kl^{3}}=0%
\text{ \ \ \ .}  \tag{8.17}
\end{equation}%
The solution of the equation can be represented as 
\begin{equation}
C_{4}(x_{i},l)=-\frac{\varepsilon _{2}e^{\frac{1}{12}}(1-const\text{ }%
e^{24k\varepsilon _{1}y})}{8k\text{ }const\text{.}}\widetilde{C}_{4}(x_{i})%
\text{ \ \ \ \ .}  \tag{8.18}
\end{equation}%
The unknown function $\widetilde{C}_{4}(x_{i})$ can be found if expression
(8.5) for $x_{1}$ is differentiated with respect to $x_{i}$. Unfortunately,
the resulting formulae will contain the expressions for $\frac{\partial y}{%
\partial x_{i}}$ and $\frac{\partial l}{\partial x_{i}}$, which are singular
when $h^{^{\prime }}(y)=0.$But if we multiply by $\frac{\partial x_{i}}{%
\partial y}$, the following differential equation will be obtained 
\begin{equation*}
\frac{\partial x_{1}}{\partial y}=\frac{\varepsilon _{1}(\varepsilon
_{2}-\varepsilon _{1})}{4}e^{2k\varepsilon _{1}y}+\frac{\varepsilon _{2}e^{%
\frac{1}{12}}}{8k\text{ }const\text{ }l^{3}}\widetilde{C}_{4}(x_{i})\frac{%
\partial l}{\partial y}-
\end{equation*}%
\begin{equation}
-\frac{\varepsilon _{2}e^{\frac{1}{12}}}{8k\text{ }const\text{ }l^{2}}\frac{%
\partial \widetilde{C}_{4}(x_{i})}{\partial y}\text{ \ \ \ \ ,}  \tag{8.19}
\end{equation}%
which is no longer singular in the limit $h^{^{\prime }}(y)=0$, because the
expressions for $\frac{\partial x_{1}}{\partial y}$ and $\frac{\partial l}{%
\partial y}$ are 
\begin{equation}
\frac{\partial x_{1}}{\partial y}=0\text{ \ \ ; \ \ }\frac{\partial l}{%
\partial y}=\varepsilon _{2}2kl(l^{2}-1)\text{ \ \ \ .}  \tag{8.20 }
\end{equation}%
Taking into account these formulaes, the solution of the differential
equation (8.19) with respect to the function $\widetilde{C}_{4}(x_{i})$ can
be found in the form 
\begin{equation}
\widetilde{C}_{4}(x_{i})=const_{2}\mid 1-const\text{ }e^{24k\varepsilon
_{1}y}\mid ^{-\frac{\varepsilon _{1}}{12}}\text{ \ .}  \tag{8.21 }
\end{equation}%
Substituting into (8.18), the final expressions for the function $%
C_{4}(x_{i},l)$ can be found and also for the coordinate $x_{1}$, which for $%
\varepsilon _{1}=\varepsilon _{2}$ and $h(y)=1$ is simply 
\begin{equation*}
x_{1}=C_{4}(x_{i}(y),l(y))=
\end{equation*}%
\begin{equation}
=-\frac{\varepsilon _{2}\text{ }const_{2}\text{ }e^{\frac{1}{12}}}{8k\text{ }%
const\text{ }}\frac{\left( 1-const\text{ }e^{24k\varepsilon _{1}y}\right) }{%
\mid 1-const\text{ }e^{24k\varepsilon _{1}y}\mid ^{\frac{\varepsilon _{1}}{12%
}}}\text{ \ \ \ \ \ .}  \tag{8.22 }
\end{equation}

\section{\protect\bigskip DISCUSSION}

It is perhaps surprising \ that the quasilinear differential equation in the
preceeding section and its solution for the length function is simpler than
the other two differential equations in sections 6 and 7, when the two cases
of "compactification+rescaling" and "rescaling+compactification" have been
investigated. Nevertheless, it is recommendable to find more solutions for
the length function $l(x)$, from where it can be seen in which other cases
the transition to $l=1$ can be performed. Also, solutions of these equations
can be found not only by the method of characteristics, but also in terms of
complicated functions of the Weierstrass elliptic function and its
derivative, following with slight modifications the algorithm in [2, 5]. As
a matter of fact, quite a broader class of equations, concerning the
Ginzburg-Landau model and $\lambda \Phi ^{4}$ scalar field models from
quantum field theory admit such solutions, representing uniformization
functions, depending on elliptic functions.

In this paper only one length scale has been taken into account. However,
there is a motivation for taking into consideration (at least)\ two
different length scales - these are the models with intersecting $D5$-
branes on $4D$ (orientifold) compactifications of type II B string theories
[15,16]. These models have two transverse to the $D5$-branes directions and
thus the string scale is lowered to the $TeV$ region.

Now let us try to implement the introduced notion of length scale \ with
respect to type II\ A string theory compactifications on a compact variety
of the form $T^{2}\times B_{4}$ [17], where several sets of $D$- branes with
one worldvolume dimention are wrapped on different cycles within a two
torus. The torus is obtained by quotienting the two dimensional flat space $%
R^{2}$ by the lattice of translations, generated by the vectors $e_{1}=(1,0)$
and $e_{2}=(0,1)$. The length of the cycle $(n,m)$ with the two compact
dimensions of size $R_{1}$ and $R_{2}$ is 
\begin{equation*}
\mid (n,m)\mid =(g_{ab}v^{a}v^{b})^{\frac{1}{2}}=
\end{equation*}%
\begin{equation}
=2\pi \sqrt{n^{2}R_{1}^{2}+m^{2}R_{2}^{2}+2nmR_{1}R_{2}cos\Theta }\text{ \ \
, }  \tag{9.1}
\end{equation}%
where $\Theta $ is the angle between the two vectors and $g_{ab}$ is the
symmetric matrix 
\begin{equation}
g_{ab}=\left( 
\begin{array}{cc}
R_{1}n & R_{1}R_{2}nmcos\Theta \\ 
R_{1}R_{2}nmcos\Theta & R_{2}m%
\end{array}%
\right) \text{ \ .}  \tag{9.2}
\end{equation}%
With respect to this two - dimensional model, let us define the
contravariant metric $\widetilde{g}^{ab}$, which here shall be identified
with the factorized product 
\begin{equation}
\widetilde{g}^{ab}:=v^{a}v^{b}\text{ \ .}  \tag{9.3}
\end{equation}%
This metric has the (constant) components 
\begin{equation}
\widetilde{g}^{11}:=n^{2}\text{ ; \ }\widetilde{g}^{12}:=nm\text{ ; \ }%
\widetilde{g}^{22}:=m^{2}\text{ \ \ .}  \tag{9.4}
\end{equation}%
In Appendix B\ the inverse contravariant metric components $g^{11}$,$g^{12}$
and $g^{22}$ and also the Christoffell components have been calculated, from
where it can be seen that they have a singularity at 
\begin{equation}
1-R_{1}R_{2}nmcos^{2}\Theta =0\text{ \ \ .}  \tag{9.5}
\end{equation}%
The scalar curvature $R$ will be singular too at the values for $\Theta $,
satisfying (9.5). However, if a new scalar curvature $\widetilde{R}$ with
the contravariant components (9.4) is defined and the equality of the two
curvatures $R$ and $\widetilde{R}$ is required, i.e. 
\begin{equation}
R=\widetilde{R}\text{ \ \ \ ,}  \tag{9.6}
\end{equation}%
then from (9.6) and the resulting algebraic equation with respect to $%
cos\Theta $ the angle $\Theta $ can be determined. The \textquotedblright
singular\textquotedblright\ value $cos^{2}\Theta =\frac{1}{R_{1}R_{2}nm}$
has to be excluded, because then (9.6) cannot be satisfied. This example
clearly shows that although the length scale is properly defined, the
singularity in the scalar curvature might not be removed by the choice of
another contravariant metric. However, it may be interesting to solve
equation (9.6) (now treated as a nonlinear differential equation), assuming
that the contravariant \textquotedblright tilda\textquotedblright\
components $\widetilde{g}^{ab}=v^{a}v^{b}=f^{a}f^{b}$ are not constant
values, but some arbitrary functions $f^{a}(R_{1},R_{2})$, and see which are
the singular and non - singular solutions.

\section*{APPENDIX\ A: AFFINE\ \ GEOMETRY\ THEOREM \ ABOUT\ \ THE\ \
EQUIAFFINE\ \ CONNECTION}

The formulation of the theorem is the following:\ if $\widetilde{\Gamma }%
_{ij}^{k}$ is another connection, not compatible with the initial one $%
\Gamma _{ij}^{k}$, then the Ricci tensor $\widetilde{R}_{ij}$ is again
symmetric and is equal to 
\begin{equation}
\widetilde{R}_{ij}=\partial _{k}\widetilde{\Gamma }_{ij}^{k}-\partial _{i}%
\widetilde{\Gamma }_{kj}^{k}+\widetilde{\Gamma }_{kl}^{k}\widetilde{\Gamma }%
_{ij}^{l}-\widetilde{\Gamma }_{ki}^{m}\widetilde{\Gamma }_{jm}^{k}\text{ \ \
\ }  \tag{A1}
\end{equation}%
if and only if the connection $\widetilde{\Gamma }_{ij}^{k}$ is an
equiaffine one. The last means that this connection for $j=k$ should be
possible to be represented as a gradient of a scalar quantity $\widetilde{e}$%
: 
\begin{equation}
\widetilde{\Gamma }_{ij}^{j}=\partial _{i}ln\widetilde{e}\text{ \ \ .} 
\tag{A2}
\end{equation}%
The equiaffine properties of the connection $\widetilde{\Gamma }_{ij}^{k}$
in the general case had been proved in a previous paper [1].

Now we shall give a verysimple proof for the concrete case $\widetilde{g}%
^{ij}=l\delta _{r}^{j}g^{ri}$, investigated in this paper . Namely, provided
that the equiaffine property (A2) is valid for the usual Christoffell
connection, after resolution of the differential equation with respect to $e$
one obtains 
\begin{equation}
e=exp\left[ C(x^{0},x^{1},..x^{i-1},x^{i+1},..,x^{n})\int \Gamma
_{ik}^{k}dx^{i}\right] \text{ \ \ .}  \tag{A3}
\end{equation}%
Then, making use of the defining equality (1.1), the \textquotedblright
tilda\textquotedblright\ connection $\widetilde{\Gamma }_{ij}^{k}$ will be
an equiaffine one if 
\begin{equation}
\widetilde{e}=exp\left[ \widetilde{C}%
(x^{0},x^{1},..x^{i-1},x^{i+1},..,x^{n})\int l\partial _{i}lnedx^{i}\right] 
\text{ \ \ .}  \tag{A4}
\end{equation}%
Substituting above the expression (A3) for $lne$ and chosing for convenience 
$\widetilde{C}=\frac{1}{C}$ (we presume that $C$ is known), one obtains the
simple expression 
\begin{equation}
\widetilde{e}=exp\left[ \int l\Gamma _{ik}^{k}dx^{i}\right] \text{ \ \ .} 
\tag{A5}
\end{equation}%
Thus, if the scalar density $\widetilde{e}$ is determined in this way, (A2)
is fulfilled, which proves that $\widetilde{\Gamma }_{ij}^{k}$ is an
equiaffine connection. Consequently, the use of formulae (A1) for the
\textquotedblright modified\textquotedblright\ Ricci tensor is fully
justified and legitimate.

\section*{APPENDIX\ B:\ CONNECTION\ AND\ \ CURVATURE\ \ COMPONENTS\ \ FOR\
THE\ TWO-DIMENSIONAL\ \ METRIC\ }

\bigskip The components $g^{11}$, $g^{22}$ and $g^{12}$, obtained after the
solution of the linear algebraic system $g_{ij}g^{jk}=\delta _{i}^{k}$ for
values of $(i,k)=(1,1)$, $(1,2)$, $(2,1)$ and $(2,2)$ are 
\begin{equation}
g^{11}=-\frac{g_{22}}{(g_{12}^{2}-g_{11}g_{22})}=\frac{1}{%
R_{1}n(1-R_{1}R_{2}nmcos^{2}\Theta )}\text{ \ ,}  \tag{B1}
\end{equation}
\begin{equation}
g^{12}=\frac{g_{12}}{(g_{12}^{2}-g_{11}g_{22})}=-\frac{cos\Theta }{%
(1-R_{1}R_{2}nmcos^{2}\Theta )}\text{ \ ,}  \tag{B2}
\end{equation}
\begin{equation}
g^{22}=-\frac{g_{11}}{(g_{12}^{2}-g_{11}g_{22})}=\frac{1}{%
R_{2}m(1-R_{1}R_{2}nmcos^{2}\Theta )}\text{ \ .}  \tag{B3}
\end{equation}

The \textquotedblright usual\textquotedblright\ Christoffell connection
components $\Gamma _{ij}^{k}$, calculated by means of the inverse
contravariant metric and the derivatives in respect to the variables $%
(R_{1},R_{2})$ are 
\begin{equation}
\Gamma _{11}^{1}=\frac{1-2R_{1}R_{2}nmcos^{2}\Theta }{%
2R_{1}(1-R_{1}R_{2}nmcos^{2}\Theta )}=g^{12}g_{12,1}+\frac{1}{2}%
(g^{11}g_{11,1}-g^{12}g_{11,2})\text{ \ ,}  \tag{B4}
\end{equation}%
\begin{equation}
\Gamma _{22}^{1}=\frac{mcos\Theta }{2(1-R_{1}R_{2}nmcos^{2}\Theta )}%
=g^{11}g_{12,2}+\frac{1}{2}(g^{12}g_{22,2}-g^{11}g_{22,1})\text{ \ ,} 
\tag{B5}
\end{equation}%
\begin{equation}
\Gamma _{11}^{2}=\frac{ncos\Theta }{2(1-R_{1}R_{2}nmcos^{2}\Theta )}%
=g^{22}g_{12,1}+\frac{1}{2}(g^{12}g_{11,1}-g^{22}g_{11,2})\text{ ,}  \tag{B6}
\end{equation}%
\begin{equation}
\Gamma _{22}^{2}=\frac{1-2R_{1}R_{2}nmcos^{2}\Theta }{%
2R_{2}(1-R_{1}R_{2}nmcos^{2}\Theta )}=g^{12}g_{12,2}+\frac{1}{2}%
(g^{22}g_{22,2}-g^{12}g_{22,1})\text{ ,}  \tag{B7}
\end{equation}%
\begin{equation}
\Gamma _{12}^{1}=\Gamma _{12}^{2}=0\text{ \ \ \ .}  \tag{B8}
\end{equation}%
The components of the \textquotedblright tilda\textquotedblright\ connection 
$\widetilde{\Gamma }_{ij}^{k}$, calculated as a linear combination of the
above connection components (B4) - (B8) are 
\begin{equation}
\widetilde{\Gamma }_{11}^{1}=(\widetilde{g}^{11}g_{11}+\widetilde{g}%
^{12}g_{12})\Gamma _{11}^{1}+(\widetilde{g}^{11}g_{12}+\widetilde{g}%
^{12}g_{22})\Gamma _{11}^{2}\text{ ,}  \tag{B9}
\end{equation}%
\begin{equation}
\widetilde{\Gamma }_{22}^{1}=(\widetilde{g}^{11}g_{11}+\widetilde{g}%
^{12}g_{12})\Gamma _{22}^{1}+(\widetilde{g}^{11}g_{12}+\widetilde{g}%
^{12}g_{22})\Gamma _{22}^{2}\text{ ,}  \tag{B10}
\end{equation}%
\begin{equation}
\widetilde{\Gamma }_{11}^{2}=(\widetilde{g}^{12}g_{11}+\widetilde{g}%
^{22}g_{12})\Gamma _{11}^{1}+(\widetilde{g}^{12}g_{12}+\widetilde{g}%
^{22}g_{22})\Gamma _{11}^{2}\text{ ,}  \tag{B11}
\end{equation}%
\begin{equation}
\widetilde{\Gamma }_{22}^{2}=(\widetilde{g}^{12}g_{11}+\widetilde{g}%
^{22}g_{12})\Gamma _{22}^{1}+(\widetilde{g}^{12}g_{12}+\widetilde{g}%
^{22}g_{22})\Gamma _{22}^{2}\text{ ,}  \tag{B12}
\end{equation}%
\begin{equation}
\widetilde{\Gamma }_{12}^{1}=\widetilde{\Gamma }_{12}^{2}=0\text{ \ \ \ .} 
\tag{B13}
\end{equation}%
The \textquotedblright non - tilda\textquotedblright\ scalar curvature $R$,
calculated by means of the formulaes (B1) - (B8) is 
\begin{equation}
R=\frac{cos\Theta \left( 2-R_{1}R_{2}mncos^{2}\Theta \right) }{%
4R_{1}R_{2}\left( 1-R_{1}R_{2}mncos^{2}\Theta \right) ^{3}}\text{ \ \ .} 
\tag{B14}
\end{equation}%
The \textquotedblright tilda\textquotedblright\ scalar curvature, calculated
by means of the connections (B9) - (B13) is 
\begin{equation*}
\widetilde{R}=\frac{m^{3}n^{3}cos\Theta (1-2R_{1}R_{2}mncos^{2}\Theta )A}{%
4(1-R_{1}R_{2}mncos^{2}\Theta )^{2}}+
\end{equation*}%
\begin{equation}
+\frac{m^{3}n^{3}(m+n^{2}R_{1}cos\Theta )(n+m^{2}R_{2}cos\Theta )B}{%
4R_{2}(1-R_{1}R_{2}mncos^{2}\Theta )^{2}}\text{ \ \ ,}  \tag{B15}
\end{equation}%
where $A$ and $B$ denote the expressions 
\begin{equation}
A:=mR_{2}(m+n^{2}R_{1}cos\Theta )^{2}+nR_{1}(n+m^{2}R_{2}cos\Theta )^{2}%
\text{ \ \ ,}  \tag{B16}
\end{equation}%
\begin{equation}
B:=-1+5mnR_{1}R_{2}cos^{2}\Theta -4R_{1}^{2}R_{2}^{2}m^{2}n^{2}cos^{4}\Theta 
\text{ \ \ .}  \tag{B17}
\end{equation}

\bigskip Now it is seen that the equality $R=\widetilde{R}$ is not satisfied
at the singular value $cos^{2}\Theta =\frac{1}{R_{1}R_{2}mn}$, although in
the formal sense it is well defined. But yet the tilda contravariant
components $\widetilde{g}^{ij}$ are well - defined, unlike $g^{ij}$.

The other values of $\Theta $ can be determined from the resulting algebraic
equation with respect to $\Theta $.

\section*{Acknowledgments}

\bigskip This paper is written in memory of Prof. Nikolai Alexandrovich
Chernikov (16. 12. 1928 - 17. 04. 2007) (BLTP, JINR, Dubna) , to whom I am
indebted for my understanding of non-euclidean (Lobachevsky) geometry.

The author is very grateful to Prof. V. V. Nesterenko (BLTP, JINR, Dubna)
and to Dr. O. P. Santillan (IAFE, Buenos Aires), Dr. N. S. Shavokhina (BLTP
\& LNP,JINR, Dubna) for valuable comments, discussions and critical remarks.

\ The author is grateful also to Dr.C. Kokorelis (Institute for Nuclear \&
Particle Physics N. C. S. R. Demokritos, Athens, Greece) and Dr. Al. Krause
(ASC, Munich) for bringing some references to my attention.

\end{document}